\documentclass{appolb}
\usepackage{epsfig}

\begin{document}
\title{Confinement and Chiral Symmetry in a Dense Matter%
}
\author{L. Ya. Glozman
\address{ Institute for Physics, Theoretical Physics Branch, University of Graz,
Universit\"atsplatz 5, A-8010 Graz, Austria}
}
\maketitle
\begin{abstract}
We discuss a possibility for existence of confining but chirally symmetric
phase at large baryon densities and low temperatures.
\end{abstract}
  
\section{General discussion}

How does the strongly interacting matter look like at
large (moderately large) baryon density and how does QCD work in this regime?
It is one of the most interesting questions for the next decade. At the same
time lattice QCD will not help us - it intrinsically relies on the Monte-Carlo 
sampling in Euclidean space-time, but at finite chemical potential the Boltzman 
weight factor becomes complex so one cannot decide which specific amplitude (path) 
is important and which is not. It is a conceptual problem and unlikely could be overcome.
We will have to rely on  empirical results and on our qualitative
understanding of QCD.

The most important and interesting question is related to interconnections
of confinement and chiral symmetry breaking and how they  both 
influence possible
phases at finite density and determine a nature of mass of a dense matter.
What will be a phase at  not high temperatures and at densities higher
than the nuclear matter density? We know a bit about the subject at the
very  large densities. At $N_c=3$ and asymptotically large chemical
potential at low temperatures the confining color-electric part of the
gluonic field is totally Debye screened by the quark-quarkhole loops and one
obtains a color-superconducting phase \cite{CS}. In this phase a matter is
a liquid and physics is dictated by the diquark Cooper pairs near the
Fermi surface. At small densities, around the nuclear matter density, the
strongly interacting matter is also a liquid. Here quarks and gluons are
confined inside nucleons, chiral symmetry is strongly broken and physics is 
mostly driven by interactions between nucleons induced by spontaneous breaking
of chiral symmetry. Whether the QCD matter at $N_c=3$ will be a liquid or a crystal
at densities above the nuclear matter densities but below the very high
densities of the color-superconductor depends on many fine details of
dynamics that is not under our control. However, it is hard to imagine that
it is a crystal, because then a phase diagram at low temperatures with
increasing density would be a sequence of the following type: liquid - crystal - liquid.
We do not know any example of this type in Nature. A natural assumption
would be that at low temperatures we always have a liquid, all the way up to very
high density. Obviously, properties of this liquid should be quite different
at different densities.
 
 This picture is drastically different from a crystalline structure that persists in the Skyrmionic
description  of a dense matter and even of a few-nucleon systems \cite{S}. The
crystalline structure in this case is entirely because of a topological nature of
the nucleon within the Skyrme model. Crystal breaks both translational and 
rotational invariances. A crystalline type structure (chiral spiral) is also
observed in solvable 1+1 dimensional models (large $N_c$ QCD in 1+1 dimensions
and Gross-Neveu model) \cite{Th}. In this case a translational invariance is
broken even for the one-nucleon solution, which is certainly not the case in the real
4-dimensional world at $N_c=3$. 

If the QCD matter is a liquid at intermediate and high densities, then,
by assumption,  both translational and 
rotational invariances are not broken. Recent lattice results for $N_c=2$ QCD \cite{Ha} support
such assumption.

Another key question is what degrees of freedom determine thermodynamical
properties of the system. In the confining mode such degrees of freedom are
color-singlet excitations of the system. Hence, in the
confining mode relevant degrees of freedom are the color-singlet baryons or
baryon-like systems near the Fermi surface. This is certainly the case at the 
nuclear matter density. A key question is then at which density (chemical potential) 
a deconfining transition occurs in the $N_c=3$ world? 

In the large $N_c$ world with quarks in the fundamental representation there
are neither dynamical quark-antiquark nor quark-quarkhole loops. Consequently
there is not Debye screening of the confining gluon propagator and a gluodynamics 
at low temperatures is
the same as in vacuum. Confinement persists up to arbitrary large 
density. In such  case it is possible to define a quarkyonic matter \cite{LP}. In short,
it is a strongly interacting matter with confinement and with a well-defined Fermi
sea of baryons or quarks. At smaller densities it should be a Fermi sea of nucleons
(so it matches with standard nuclear matter), while at higher densities, when  nucleons 
are in a strong overlap, a quark Fermi surface should be formed. While a quark Fermi sea
is formed, the system is still with confinement and excitation modes are of the color-singlet
hadronic type. 

In the large $N_c$ 't Hooft limit such a matter persists at low temperatures 
up to arbitrary large densities. At
which densities in the real  $N_c=3$ world will we have a deconfining transition (which could be a very
smooth crossover) to a quark matter with uncorrelated single quark  excitations? 
Lattice results for the $N_c=2$ suggest that such
a transition could occur at densities of the order 100*nuclear matter density \cite{Ha}.
If correct, then it would imply that at all densities relevant to future experiments
and astrophysics we will have a dense quarkyonic matter with confinement.

A very interesting question is what happens with chiral symmetry dynamical
breaking in such a dense matter with confinement. If it is a liquid with
unbroken translational and rotational invariances one  expects that due
to Pauli blocking of the quark levels (required for the very existence of the
quark-antiquark condensate of the vacuum) there will happen a chiral restoration
phase transition and we will obtain a confining but chirally symmetric
subphase within the quarkyonic matter \cite{G1,G2,G3}. In such a phase a bulk
mass of the system is generated  via the chirally-invariant dynamics.

We cannot solve QCD and such a conclusion cannot be obtained from QCD itself.
At best what we can do at the present stage is to answer a question whether it
is possible or not in principle. If yes, then a key question is: {\it How} could it
happen and which physics is behind such unusual situation?
Consequently we need a  confining and chirally symmetric model. We cannot expect any realistic numbers
from such a model, because this model would be at best a great oversimplification
of real QCD. However, still such  illustrative model is important because
it gives us the insight into physics and because
it was believed for many years that a confining phase with vanishing quark-antiquark
condensate of the vacuum is impossible. Definitive requirements for the model are that
it must be confining, chirally symmetric, provide dynamical chiral symmetry breaking
in the vacuum. Obviously, the issue of a matter with confinement and restored
chiral symmetry cannot be formulated and studied with the Nambu and Jona-Lasinio model and its 
variants which  has been used sofar for study of the phase diagram of the strongly interacting matter.

\section{The model}

The model which we will use as a laboratory to get the insight is a generalization
of the 't Hooft model, that is QCD in large $N_c$ limit in 1+1 dimensions. In the latter
case the only gluonic interaction is a linear confining instantaneous Coulomb potential
between quarks. Solving a gap equation one obtains a quark Green function with dynamical
mass. Given this quark Green function it is possible to calculate exactly a meson
spectrum, a quark condensate, etc. Consider a straightforward generalization,
i.e., a model with a linear instantaneous
Coulomb-like potential in 3+1 dimensions \cite{Ya,Ad}. All other possible gluonic
interactions are neglected. A very important aspect of this model is that it exhibits
the effective chiral restoration in hadrons with large $J$  \cite{WG}. This means that their
mass comes  not from the quark condensate of the vacuum, but mostly from the manifestly
chiral-invariant dynamics. A chiral symmetry breaking in the vacuum  is only 
a tiny perturbation to the chiral-invariant mass of these high-spin hadrons. 
It explicitly demonstrates that it
is possible to construct  hadrons such that their mass origin is not the quark condensate
of the vacuum. If so, it is clear apriori that there are good chances to obtain confining
but chirally symmetric matter at low temperatures within this model \cite{G1,G2,G3}.

A key point is that the quark Green function (that is  a solution of a gap
equation in a vacuum) contains not only a chiral symmetry breaking part $A_p$,
but also a manifestly chirally symmetric part $B_p$: 

\begin{equation}
\Sigma(\vec p) =A_p +(\vec{\gamma}\cdot\hat{\vec{p}})[B_p-p].
\label{SE} 
\end{equation}

A linear potential requires the infrared regularization. Otherwise all loop
integrals are infrared divergent. All observable color-singlet quantities
are finite and well defined in the infrared limit (i.e., when the infrared cutoff approaches
zero). These are hadron masses, the quark condensate, etc. In contrast,  all color-nonsinglet quantities 
are divergent. E.g., single quarks have infinite energy and consequently are removed from the spectrum.  
This is a manifestation of confinement within this simple model.

Assume now that we have a dense baryon matter with a well defined quark Fermi sea and the quark 
Fermi momentum is $P_f$. At the same time the interquark linear potential is not
yet screened (in the  large $N_c$ limit it is not screened at any density). In order to 
understand what happens with  chiral symmetry
we have to solve a gap equation for a probe quark, see the left figure below. It is also assumed that the translational and
rotational invariances are not broken according to  arguments of the previous section. All intermediate 
quark levels below $P_f$ are Paili blocked and do not contribute to the gap equation. Consequently, at sufficiently
large $P_f$ a chiral restoration phase transition happens, see the right figure.

\begin{center} \includegraphics[height=3cm]{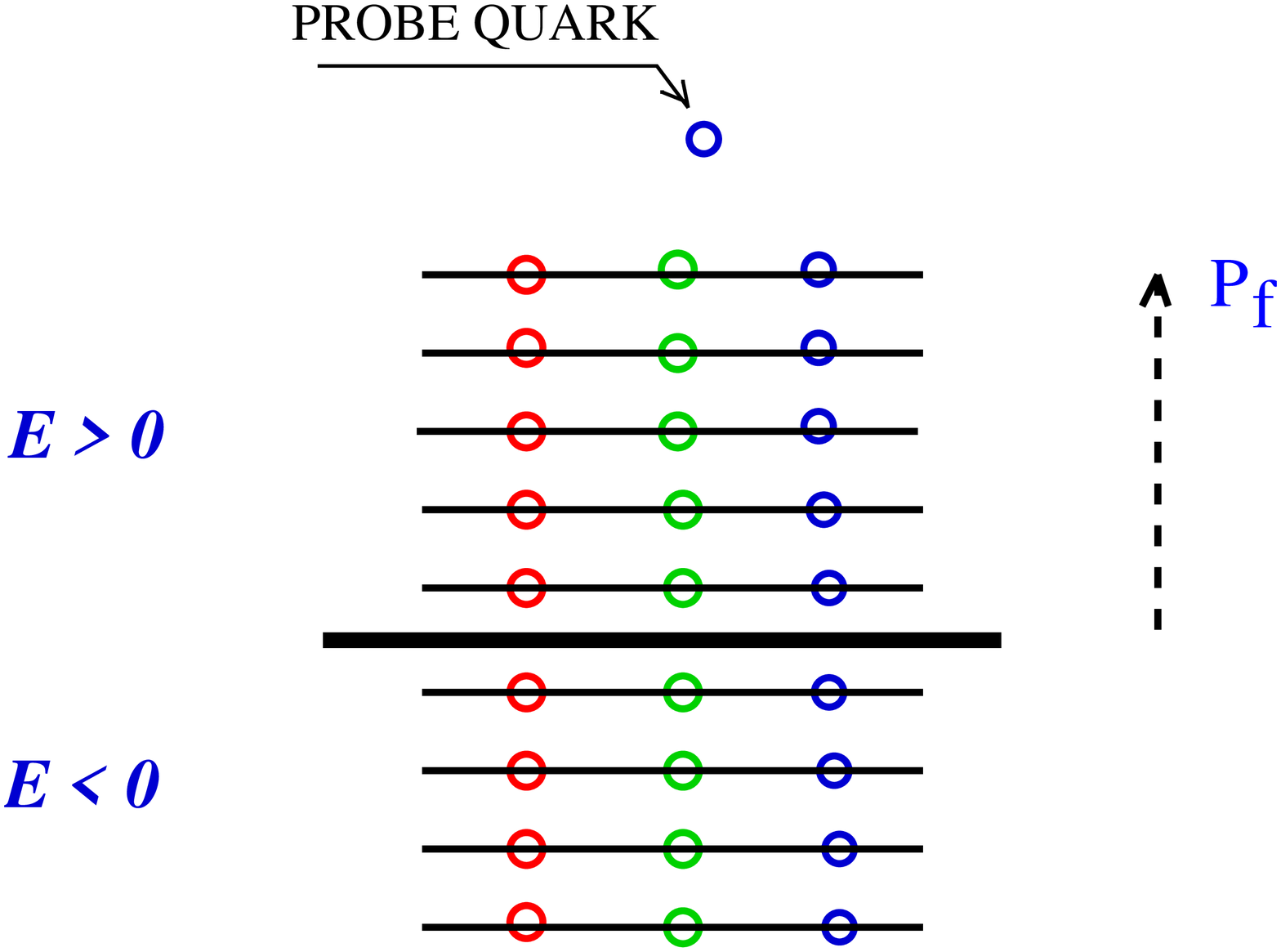}
\includegraphics[height=3cm,clip=]{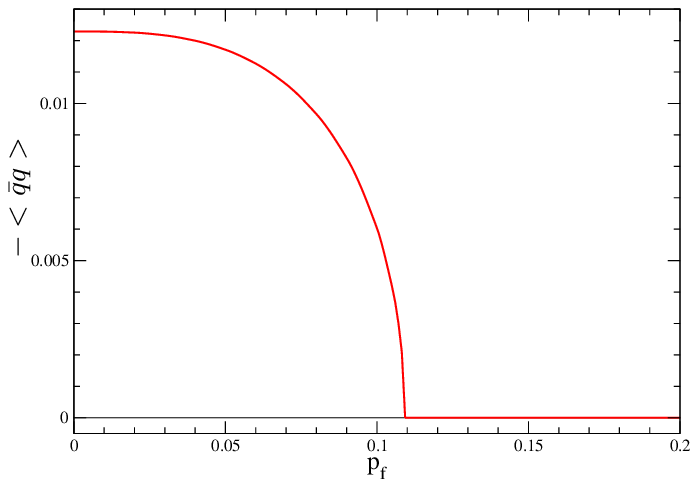}
\end{center}

The chiral symmetry is restored like in the NJL model, because there is not available phase space
in the gap equation to create a nontrivial solution with broken chiral symmetry. This required
phase space is removed by the Pauli blocking of  levels with positive energy. The standard 
quark-antiquark condensate of the vacuum vanishes.

As a consequence, above the chiral restoration phase transition the chiral symmetry breaking
part of the quark self-energy identically vanishes, $A_p=0$. However, there is also a chirally
symmetric part of the quark self-energy, $B_p$. It does not vanish and is still infrared-divergent,
like in vacuum. This means that even in the chirally restored regime a single quark energy is infinite
and a single quark is removed from the spectrum. This infrared divergence is exactly canceled, however,
in any color-singlet excitations of baryonic or mesonic type. Consequently, a spectrum of excitations
consists of a complete set of all possible chiral multiplets. Energy of these excitations is a finite
and well-defined quantity. The mass of this confining matter is chirally symmetric
and comes  from the chiral-invariant dynamics.

Such a confined phase with restored chiral symmetry can be viewed as a system of chirally
symmetric baryons that are in a strong overlap, see a cartoon below. Confining gluonic
fields are not screened, but quarks can move not only within each individual baryon,
but also within the matter by hoping from one baryon to another. Such a motion is a simple consequence of
Pauli principle. So one cannot say to which specific baryon a given quark belongs. (Actually 
even in a deuteron, which is certainly in a confining mode, one cannot say to which specific
nucleon a given quark belongs: The valence quarks are transported from one nucleon to another
by the meson-exchange force.). However, it would be a mistake to consider these quarks as free
particles. They are colored and are subject to strong gluonic confining fields. Their dispersion
law is a complicated one, by far not as for free particles. This picture is different from the
naive perlocation picture, where one thinks that due to perlocation of baryons the quarks are
free.

\begin{center}
\includegraphics[height=1.5cm]{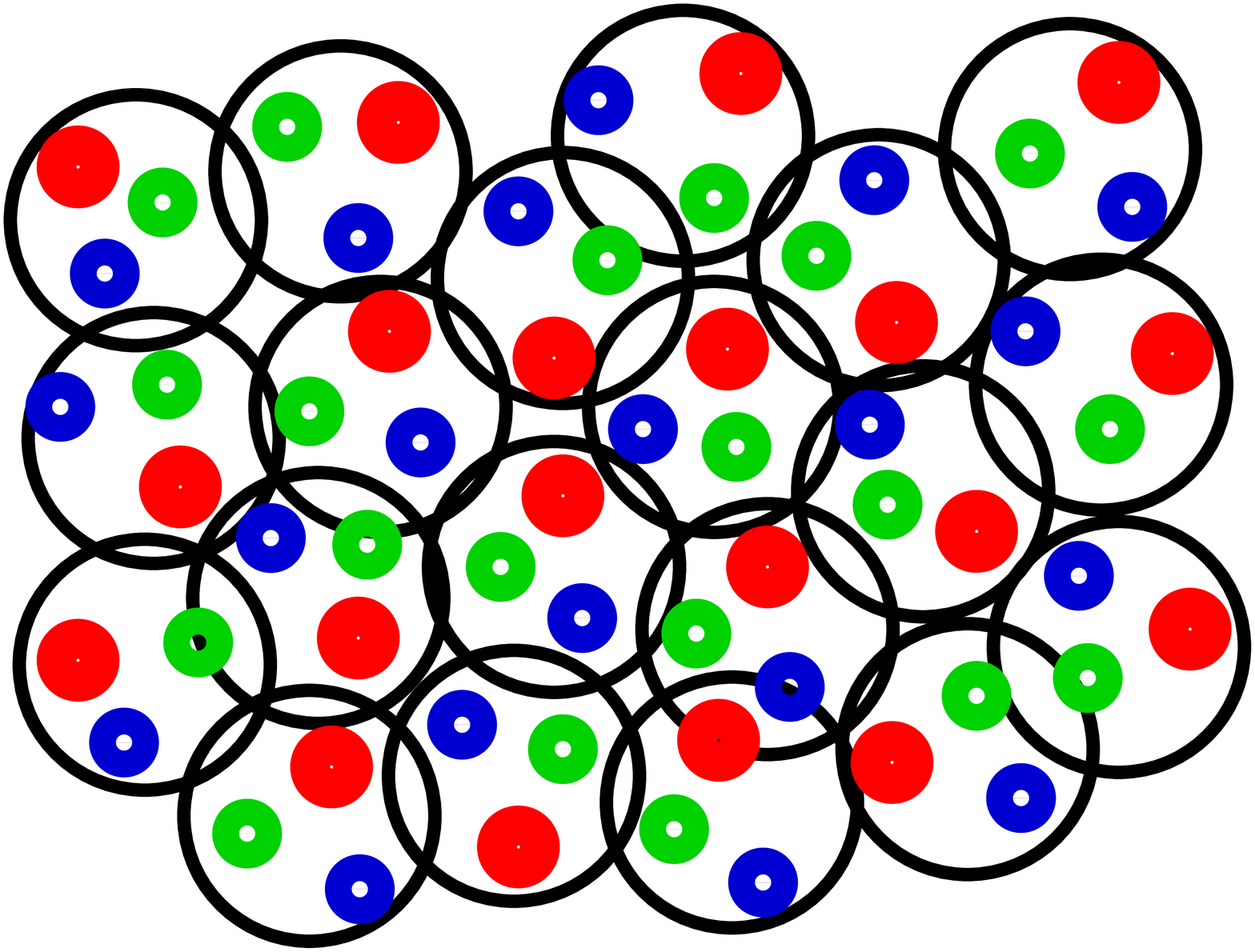}
\end{center}

It should be emphasized that existence of chirally symmetric hadrons at large
density is not prohibited and the Casher argument \cite{C} can
be easily bypassed in this situation \cite{G3}.

An interesting question is what happens near the Fermi surface of such a dense
confining matter with vanishing quark-antiquark condensate. There could be some
surface phenomena like chiral density waves  \cite{Ru}. These chiral density
waves have been derived so far as an instability of the quark Fermi sea
(with free unconfined quarks) due to a gluon-exchange force between
quarks and quarkholes with large momenta near the Fermi surface. It is by far
not clear whether it will happen or not in a system with confinement.

Below we show a possible phase diagram that incorporates  a phase
with confinement (or its remnants) but with vanishing quark-antiquark condensate
as a phase at low temperatures just between a dense nuclear matter and a color-superconducting
phase. 

\begin{center}
\includegraphics[height=4cm]{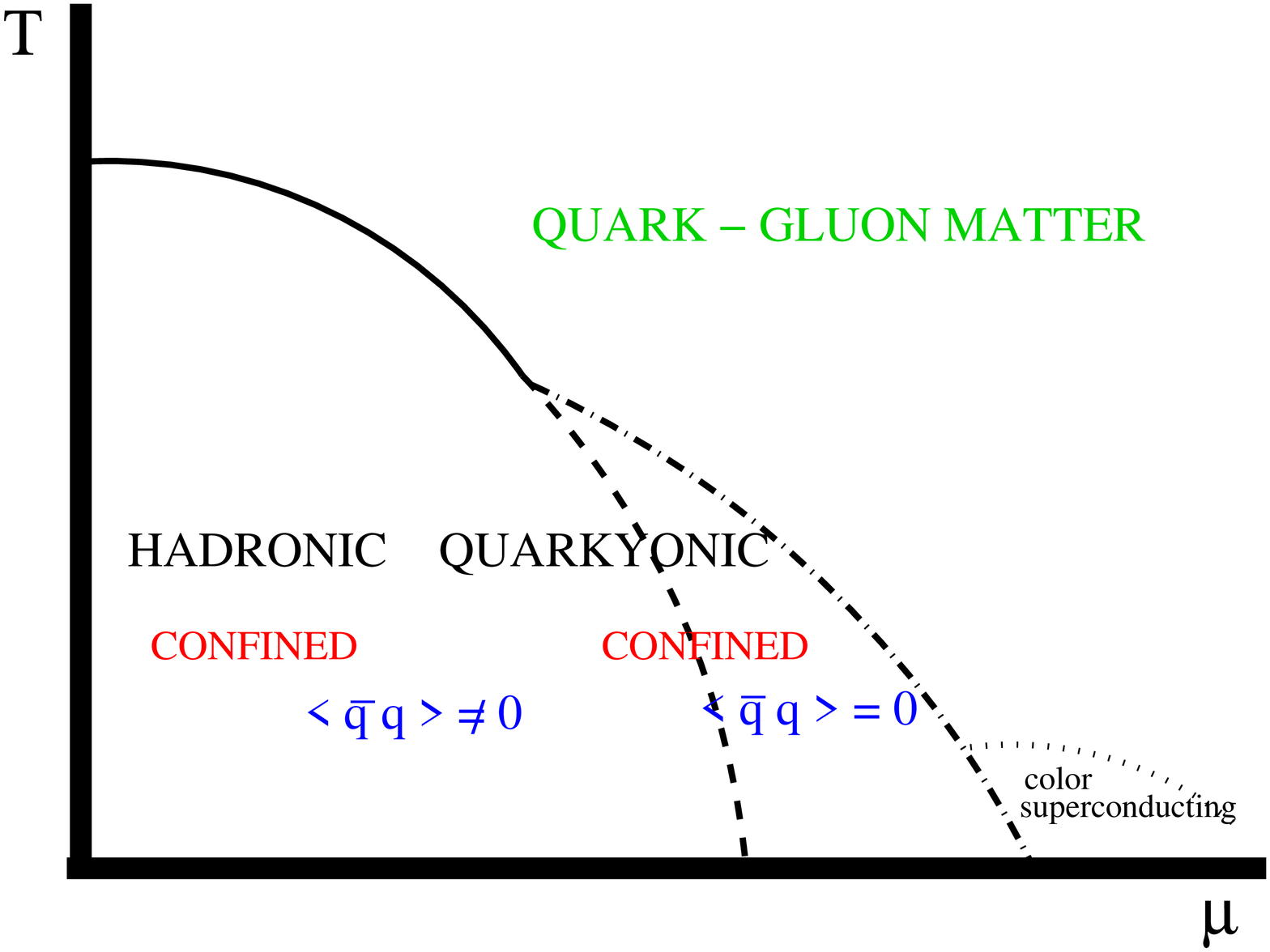}
\end{center}

\medskip
{\bf Acknowledgements}
Support of the Austrian Science
Fund through the grant P21970-N16 is acknowledged.

\end{document}